\documentclass[10pt, conference, compsocconf]{IEEEtran}
\ifCLASSINFOpdf
\else
\fi

\usepackage{mathtools}
\usepackage{amssymb}
\usepackage{tabularx}
\usepackage[makeroom]{cancel}
\usepackage{comment}
\usepackage{color}

\usepackage{hyperref}
\usepackage{subfigure}

\usepackage{xspace}
\newcommand{\semmeddb}{\textsf{SemMedDB}\xspace}
\newcommand{\medline}{\textsf{MEDLINE}\xspace}

\newcommand{\umls}{\textsf{UMLS}\xspace}
\newcommand{\ftext}{\textsf{FastText}\xspace}
\newcommand{\sysname}{\textsf{MOLIERE}\xspace}
\newcommand{\moliere}{\textsf{MOLIERE}\xspace}
\newcommand{\tmine}{\textsf{ToPMine}\xspace}

\newcommand{\brainsc}{\textsf{brainSCANr}\xspace}
\newcommand{\discon}{\textsf{DiseaseConnect}\xspace}
\newcommand{\arrowsmith}{\textsf{Arrowsmith}\xspace}

\usepackage{changes}
\definechangesauthor[color=red]{is}
\definechangesauthor[color=blue]{js}
\definechangesauthor[color=green]{ac}
\definechangesauthor[color=purple]{ah}

\usepackage{todonotes}
\makeatletter
\setremarkmarkup{\todo[color=Changes@Color#1!20,size=\scriptsize]{#1: #2}}
\makeatother

\usepackage{graphicx}
\usepackage{calc}
\newlength{\depthofsumsign}
\newlength{\totalheightofsumsign}
\newlength{\heightanddepthofargument}

\makeatletter
\newcommand*{\DivideLengths}[2]{%
  \strip@pt\dimexpr\number\numexpr\number\dimexpr#1\relax*65536/\number\dimexpr#2\relax\relax sp\relax
}
\makeatother

\graphicspath{{img/}}

\usepackage{algpseudocode}
\usepackage{algorithm}

\algblock{ParFor}{EndParFor}
\algnewcommand\algorithmicparfor{\textbf{for}}
\algnewcommand\algorithmicpardo{\textbf{pardo}}
\algnewcommand\algorithmicendparfor{\textbf{end\ for}}
\algrenewtext{ParFor}[1]{\algorithmicparfor\ #1\ \algorithmicpardo}
\algrenewtext{EndParFor}{\algorithmicendparfor}

\definecolor{purple}{HTML}{800080}
\definecolor{blue}{HTML}{0000ff}
\definecolor{green}{HTML}{008000}
\definecolor{cyan}{HTML}{a4ffff}
\definecolor{orange}{HTML}{ffa500}
\definecolor{red}{HTML}{ff0000}

\begin{document}
\title{Are Abstracts Enough for Hypothesis Generation?}


\author{\IEEEauthorblockN{
Justin Sybrandt,
Angelo Carrabba,
Alexander Herzog,
Ilya Safro
}
\IEEEauthorblockA{
Clemson University\\
School of Computing\\
Clemson, USA\\
\{jsybran, acarrab, aherzog, isafro\}@clemson.edu}
}


%Telephone: (800) 555--1212, Fax: (888) 555--1212}


\maketitle

\begin{abstract}

The potential for automatic hypothesis generation (HG) systems to improve research productivity keeps pace with the growing set of publicly available scientific information.
But as data becomes easier to acquire, we must understand the effect different textual data sources have on our resulting hypotheses.
Are abstracts enough for HG, or does it need full-text papers?
How many papers does an HG system need to make valuable predictions?
How sensitive is a general-purpose HG system to hyperparameter values or input quality?
What effect does corpus size and document length have on HG results?
To answer these questions we train multiple versions of knowledge network-based HG system, \moliere, on varying corpora in order to compare challenges and trade offs in terms of result quality and computational requirements.
\moliere generalizes main principles of similar knowledge network-based HG systems and reinforces them with topic modeling components. The corpora include the abstract and full-text versions of PubMed Central, as well as iterative halves of MEDLINE, which allows us to compare the effect document length and count has on the results.
We find that, quantitatively, corpora with a higher median document length result in marginally higher quality results, yet require substantially longer to process.
However, qualitatively, full-length papers introduce a significant number of intruder terms to the resulting topics, which decreases human interpretability.
Additionally, we find that the effect of document length is greater than that of document count,  even if both sets contain only paper abstracts.\\
Reproducibility: Our code and data are available online at  {\color{blue} \href{http://sybrandt.com/2018/abstracts}{sybrandt.com/2018/abstracts}}.
\end{abstract}

% \fi
\IEEEpeerreviewmaketitle

\section{Introduction}\label{sec:introduction}

While the driving pace of research accelerates~\cite{larsen2010rate,van2014global}, computer-aided methods become increasingly more important for improving scientific productivity.
This is especially apparent in medicine and life sciences --- the National Institute of Health introduced 1.1 million papers to \medline in 2017 alone.
Hypothesis Generation (HG)~\cite{spangler2015accelerating} is the process of finding unknown-yet-useful connections from the set of publicly available information.
Usually, this involves a combination of text processing, data mining, and graph-based approaches.

When scientists miss cross-cutting connections, they leave behind \emph{undiscovered public knowledge}~\cite{swanson1986undiscovered}, which many aim to detect through Hypothesis Generation (also called Literature-Based Discovery) Systems~\cite{bruza2008literature,spangler2015accelerating}.
Early attempts find important connections from the co-occurrences of keywords across paper titles~\cite{smalheiser1996linking}, while more advanced methods, such as recommender systems~\cite{spangler2014automated} and topic modeling~\cite{sybrandt2017moliere}, rely on abstracts and preprocessed longer documents such as full-text papers in IBM Watson Drug Discovery system.
No matter the method, every system is primarily dependent on text, yet to the best of our knowledge no one has directly and systematically addressed the effect corpus quality, size, and document length has on the quality of knowledge network-based HG systems.

Because of huge practical importance of HG systems for accelerating biomedical discovery, there are many controversial arguments on the need of full-text papers in the scientific community~\cite{Blair1985evaluation,Schuemie2004,shah2003information,Westergaard2017}.
However, in the vast majority of studies, this issue is raised with respect to traditional \emph{information retrieval} (IR)  and \emph{data mining} tasks and systems, which usually do not substitute HG.
Clearly, full-texts are more beneficial for IR as they contain more information, but does the same hold for HG?  

\noindent{\bf Our Contribution:}
We explore the effect corpus size and document length have on knowledge network-based HG systems, primarily by comparing their performance with full-text papers against abstracts.
Our experimental studies are based on the HG system \moliere~\cite{sybrandt2017moliere} that extends the basic principals of knowledge discovery networks introduced in earlier works~\cite{smalheiser1998using,spangler2015accelerating,wang2011finding}.
This centers around two major studies: the first comparing the performance of our system trained on abstract and full-text versions of the same document set, the second comparing the performance of iterative halves of a large abstract set.
Our results, while experimentally focused on \moliere, have important implications to other similar systems~\cite{spangler2014automated,voytek2012automated,wang2011finding}.
%\TODO{Throw in a couple more citations}

We evaluate our results in terms of quality, using the hypothesis ranking techniques developed in~\cite{sybrandt2018validation}, and discuss practical challenges in terms of memory consumption, runtime, and interpretability.
We find that corpora with a higher median document length perform better than those with shorter documents and that this effect can be more substantial than simply adding more documents.
Most importantly, when comparing a corpus of full-text documents against a corpus of the abstracts of those same documents, we notice a \emph{marginal} improvement in quality (if at all), yet a \textbf{45$\times$} increase in runtime from 100 seconds to 75 minutes.

To perform our evaluation, we create multiple \emph{instances} of our HG system.
By this we mean that we perform our entire knowledge network construction process, starting from raw documents and ending at a large knowledge network \cite{sybrandt2017moliere}, independently for each corpus version.
We start our study with data from \medline as well as PubMed Central (PMC).
The former contains over 24 million abstracts dating back to the late 1800's, while the latter contains 4 million full-text documents (only 1.7M in XML) and started in the year 2000.
Using PMC we explore the effect document length has on HG systems by training two instances of \moliere on the abstract and full-text versions of the same corpus.
With \medline, we evaluate the effect of corpus size using five instances of \moliere trained on repeated halves of the data set.

%\replaced[id=js]{
Our validation compares instances by their ability to distinguish published from noise connections based on their resulting hypotheses, given that the instance has no available information regarding either.
%}{
%Instances that score highly in our evaluation are better able to distinguish published connections (which correspond to  generated hypotheses in our system) from noise given a training set that contains neither \added[id=is]{and no other published information about them}.
%}
This begins by selecting a \emph{cut-year} --- we choose 2015 --- and filtering our data sources to only include information that was available prior to it.
We extract recently published connections from \semmeddb, a database of \medline \emph{predicates} (subject-verb-object structures), by identifying those predicates that first occur after the cut year~\cite{Kilicoglu2012}.
We additionally create an equal number of randomly sampled connections that do not exist within \semmeddb.
By generating hypotheses for all of these connections, and ranking their results with regard to a number of metrics, we plot ROC curves that describe the quality of our system.

\section{Background: Literature-Based HG}\label{sec:background}
%\subsection{Literature-Based Hypothesis Generation}
Swanson first introduced Hypothesis Generation (HG) and his ABC model for knowledge discovery~\cite{swanson1986fish}.
He found a connection between Raynaud's syndrome (A) and fish oil (C) through their connection with blood viscosity (B).
Although Swanson's early work managed to extract these ideas using only the titles of \medline articles, recent systems, such as \brainsc~\cite{voytek2012automated}, \discon~\cite{Liu2014}, and \sysname~\cite{sybrandt2017moliere}, use modern text-mining technologies to identify latent features from abstracts in order to better extract semantic information.
These systems use abstracts for two reasons.
First, abstracts are more easily available than full-text data.
For example, \medline contains 24 million abstracts, while only 4 million full-text documents are available through PubMed Central (and most are not available in XML).
Second, there is conventional understanding that abstracts contain effective summaries of key findings~\cite{dos1996textual}, which means they have a better signal-to-noise ratio than full-text documents, which often contain textual information that is less relevant for the HG-task (e.g., references to figures, a detailed description of experimental conditions, inappropriate background).
However, the latter has not been systematically tested in the literature. 

We do, however, observe at least one commercial system that uses full-text documents.
Watson for Drug Discovery~\cite{spangler2014automated} includes a sophisticated entity extraction and ontology creation pipeline that allows it to overcome the typical signal-to-noise challenges present in these longer texts.
Additionally, the Watson discovery methods, such as co-occurrence networks and recommender systems, function on top of these pre-processed results, which means that Watson does not need to process full papers while performing individual queries.
However, we are limited in our comparison because Watson, as well as most other HG systems, are proprietary or closed-source and not available for a systematic comparison.
In Section~\ref{sec:tradeoffs} we explore the tradeoffs present between these choices of methods.

\subsection{Abstract versus Full-Text Comparisons}
Previous studies that compare abstracts and full-text
papers have done so for the purpose of information retrieval and pattern discovery in data mining (IR/DM). While IR/DM's goal is to extract known information (including finding patterns) in (un)structured data \cite{manning2008introduction}, HG's goal is to propose novel hypotheses and discover unknown information (not necessarily represented as a pattern). With this distinction in mind, it is clear
that full-text documents, by nature of their length,
contain more retrievable information than abstracts. 

Shah et al. \cite{shah2003information} perform keyword extraction from 104 articles published in Nature Genetics, showing that the full text of an article can contain as many as four times more relevant keywords than its abstract.
Schuemie et al. \cite{Schuemie2004} extract keywords from around 4,000 biomedical articles.
They similarly find that full-texts include substantially more information than abstracts, leading to a greater number of identified keywords.
Westergaard et al. \cite{Westergaard2017} confirm this finding in the context of named entity recognition (protein--protein, disease--gene, and protein subcellular associations) from 15 million biomedical full-text articles.

While the above studies show that more text is better for information extraction, they also show that there is significant heterogeneity in information density between different sections of an article. Both Shah et al. \cite{shah2003information} and Schuemie et al.  \cite{Schuemie2004} find that the information density (i.e., the ratio of relevant to irrelevant keywords) is highest in the abstract.
Given that full-text articles are more difficult to obtain, restricting the analysis to abstracts can be a sensible choice (given 24M abstracts are available through \medline, but only about 4M full-text articles are available through PMC).
Further, using full-text articles always requires significant efforts in additional text preprocessing, such as parsing parenthesized sentences or extracting text in footnotes.

Blair et al. \cite{Blair1985evaluation} note the limitations in comprehending full-texts --- longer documents typically mention many different concepts.
For instance, in our later results, we notice that many full-text documents contain significant information related to experimental procedure, which may obfuscate more relevant information regarding conclusions of new findings.
This added ``noise'' can decrease the quality of an analysis, depending on which metric is deemed most important.
Sinclair and Webber \cite{sinclair2004classification}, for example, perform Gene Ontology (GO) code classification on 1,000 articles.
Their results show that classification on full-text articles has the highest recall but lowest precision, while the opposite was true when only titles and abstracts were used.

Outside the domain of biomedical literature research, there are similarly mixed results on the question whether more text is necessarily better.
In an analysis of data from the online social network Twitter, Conover et al. \cite{conover2011predicting} find that a classifier trained on hashtags (i.e., user-selected keywords attached to a message) outperforms a classifier trained on the full text of tweets for the purpose of predicting users' political alignment.
They argue that this result is due to a better signal-to-noise ratio of keywords compared to full-text messages. 

Syed and Spruit \cite{syed2017full} apply LDA topic modeling~\cite{blei2003latent} to full-text articles and abstracts from the domain of fisheries and aquatic sciences.
Comparing the quality of estimated topics (both statistically and through human expert coders), they find that full text produces more high-quality topics than abstracts, but only when estimated on a small data set with 4,417 articles from a single journal.
On a larger data set with around 15,000 articles from 12 journals, both full text and abstracts produce similarly good results.   

To summarize, previous work has found that more text is generally better for IR/DM tasks, but many applications suffer when trained with full texts because a longer length comes with a reduced signal-to-noise ratio, even for IR/DM \cite{Cohen2010}. Given that full-text documents are much harder to acquire and require more computational resources to process, it is important to quantify these trade-offs in the context of prediction in HG.

\section{Methodology}\label{sec:methodology}

In order to understand the effect corpus size and document length has on knowledge-network-based HG systems, we train multiple instances of \moliere using data from both PubMed Central (PMC) as well as \medline.
For practical purposes, we limit our discussion to this system, but note our results have further-reaching implications.
In this section, we provide an overview of these data sources, outline our training and validation procedure, and explain the quantitative and qualitative results we collect.

\subsection{\moliere Pipeline Background}

\begin{figure*}
\centering
\subfigure[Network Construction Pipeline]{\label{fig:net_construction_pipeline}
\includegraphics[width=\textwidth]{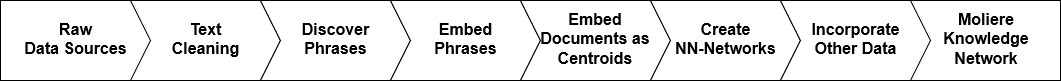}
}
\\
\subfigure[Query Processing Pipeline]{\label{fig:query_pipeline}
\includegraphics[width=.75\textwidth]{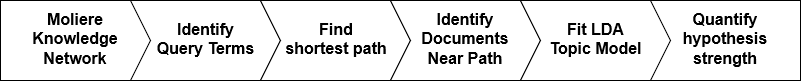}
}
\caption{Above depicts the network construction and query pipeline. First, input from raw data sources is tokenized into meaningful n-grams, then embedded, and used with other features and sources to create a nearest-neighbors network. Once the network is constructed, the query process details how we use shortest paths to identify relevant abstracts on which we generate LDA topic models.}
\end{figure*}

The process of generating fruitful hypotheses via \moliere begins with textual data sources.
In this work, we will focus on the titles and abstracts provided by \medline, or the plain-text releases of full-text papers provided by PubMed Central as our input data, but it is useful to keep in mind that \moliere is intended to work well given various input sources.
From there, we leverage recent phrase mining tools, such as \tmine\cite{el2014scalable} or AutoPhrase\cite{shang2017automated}, to segment our raw text into more easily interpretable n-grams.
We find that this step is crucial to making our downstream model output human understandable.
From there we run \ftext\cite{joulin2016bag}, a recent advance in the save vein as word2vec\cite{mikolov2013efficient}, to embed each n-gram into a 500-dimensional vector space.
This process allows us to mathematically describe the semantic similarities between our terms through simple metrics such as $L_2$ norm or cosine similarity.
We then project each document into the space by taking a weighted average of each n-grams embedding with respect to that terms TF-IDF score.
Finally, we create a nearest-neighbors network within the abstract set, and separately within the n-gram set.
Links between these sets derive from the TF-IDF scores between abstracts and n-grams.
In addition, we introduce UMLS terms, codified medical entities with known links between them, as a "backbone" to the overall network.
A diagram describing this process is shown in Figure~\ref{fig:net_construction_pipeline}.

To query this network for hypotheses, we begin with two nodes of interest, $a$ and $c$.
Typically, these are either keywords or UMLS terms.
We identify a region of the overall network containing both keywords, and run Dijkstra algorithm  within that region to quickly find a shortest-path connecting both terms.
This path, at a high level, represents a series of terms and documents that ought to outline the relationship between $a$ and $c$, but in practical cases, this path alone is not sufficient for a human scientist to form a useful hypothesis.
Therefore, we increase the amount of relevant information by taking a large set of nearby documents, typically on the order of 5-15 thousand, that are first or second-degree neighbors to the path.
This collection of nearby papers represents a sizable portion of related research, which more likely describes the nature of an $a$-$c$ relationship.
We use LDA topic models to uncover the structure of this document subset, which offers some initial insights though the clustering of interesting terms.
The process of creating these topics from relevant document sets is diagrammed in Figure~\ref{fig:query_pipeline}.

More recently, we proposed a number of metrics to evaluate $a$-$c$ relevance in~\cite{sybrandt2018validation}.
This work describes how a number of embedding-based relationships, further summarized in the following section, quantify the fruitfulness of an individual query.
While we use these metrics to validate our approach in the previously mentioned work, we leverage them here for the purpose of a numerical comparison between different data sources.

\subsection{Metrics for Hypothesis Ranking}

Many have noted key challenges that surround evaluating hypothesis generation systems~\cite{bruza2008literature}.
Because these systems attempt to locate novel research directions, unknown to even those constructing the system itself, it is difficult to distinguish a proposed hypothesis that is incorrect verses one that is true yet unintuitive.
Due to this conceptual limitation, many projects validate their system by simply rediscovering a handful of ``gold-standard'' connections~\cite{weeber2001using,Srinivasan2004,hu2005semantic,pratt2003litlinker}.
Some few projects show their utility beyond the gold-standard by incorporating expert analysis and experiments~\cite{swanson1986fish,bakkar2018artificial,sybrandt2018validation}.
While these results are important to show real-world application areas for hypothesis generation, lab work is time consuming, expensive, and clearly does not scale for large validation sets.
For these reasons, in~\cite{sybrandt2018validation}, we present a number of metrics that estimate the potential of an automatically generated hypothesis.
In that work we demonstrate the usefulness of these metrics to identify recent fruitful hypotheses given  historical training data.
Additionally, we follow the recommendations of our metrics to identify new gene treatment targets for HIV associated neurocognitive disorder.
In this work we use these same metrics to numerically compare the performance of a hypothesis generation system trained on abstracts against the same system trained on full text versions of the same papers.

Our metrics, which are summarized below, are predicated upon known properties of word embeddings.
Mikolov et al. in~\cite{mikolov2013distributed} demonstrate that their word embeddings, which were trained on 6 billion news articles from the Google News corpus, capture a latent space with meaningful distances.
For instance, the distance between the vectors for ``man'' and ``woman'' is similar to that between ``king'' and ``queen''.
This gender-encoding distance is similarly seen for other male-female relationships across the English language, which is also observed in country-capital relationships as well as that of verb tense.
Furthermore, similar words are grouped by their semantic meaning.
We observe this property in our own embeddings trained in our previous work on over 25 million \medline abstracts.

From these observations we derive the following metrics.
Here, $a$ and $c$ are two terms of a proposed hypothesis, the plausibility of which we would like to estimate.
$T$ is an LDA topic model generated from a subset of papers relevant to $a$ and $c$, and $T_i \in T$ is a single topic.
Additionally, $\epsilon(x)$ is an embedding function that maps a term or a topic into an embedding space with the previously described properties.
In the case of $\epsilon(T_i)$ we simply calculate a weighted centroid for topic $T_i$.

The simplest metric, \textsc{L$_2$}, is simply the norm of $\epsilon(a)-\epsilon(c)$.
In our previous work we also explored the cosine similarity of our term vectors, but $L_2$ was our higher performer.
Next in complexity is \textsc{CentrL$_2$} that captures the distance between the $\epsilon(a), \epsilon(c)$ midpoint from the topic model.
We observe that for a hypothesis to be supported, at least one LDA cluster ought to center between the search terms.
Then, \textsc{TopicPerWord} relaxes the assumption that topics are best represented as a centroid, and instead treats them as a weighed point cloud.
Therefore, we average the distance between each $\epsilon(x) \forall x\in T_i$, and $(\epsilon(a)+\epsilon(c))/2$, weighting them by $P(x|T_i)$.
\textsc{TopicCorr} calculates the correlation between all topics in a topic model with respect to both $a$ and $c$.
Put plainly, if a topic is close to $a$, is it also likely to be close to $c$?
Next, to calculate \textsc{TopicWalkBetweenness} we generate a nearest-neighbors network containing  $\epsilon(T_i) \forall T_i \in T$ as well as $\epsilon(a)$ and $\epsilon(c)$.
We observe that plausible hypotheses have a higher connectivity within this network, which we calculate by first finding a short path from $a$ to $c$ across $T_i$, and then calculating the average betweenness of the nodes appearing along this path.
Finally, in order to weight the heuristics present in each of the previously described metrics, we fit a polynomial based on our set of proposed hypothesis.
This results in the best-performing metric of \textsc{PolyMulti}.

\subsection{Training Corpora}

In order to understand the effect of different dataset features on an HG system, we identify corpora that differ in terms size and document length.
These data sets, outlined in Table~\ref{tab:corporaDetails}, include the PMC set of abstracts, PMC full-text, and five iterative halves of \medline.
We download each data source as XML from PMC, and apply a series of preprocessing steps, described below.
We note that while PMC contains 4 million full-text articles, a substantial number either do not supply an abstract, or are not available as XML.
While other groups have found success parsing PDF documents~\cite{Westergaard2017}, we note that future journals contributing to PMC must supply XML, and that parsing PDFs introduces a level of complexity that extends beyond the scope of this work.

We apply AutoPhrase~\cite{shang2017automated}, porter stemming, and then stopword removal to clean our text.
Our stopword list comes from \arrowsmith's list.\footnote{\url{http://arrowsmith.psych.uic.edu/arrowsmith\_uic/data/stopwords\_pubmed}}
As a result, we can identify meaningful n-grams within our text that make our results more interpretable and robust.

Because we cannot experimentally increase the size of our data sets, we instead take iterative halves of \medline until it falls below one million abstracts.
We do so with random sampling without replacement, and we note that the smaller samples are contained within the larger corpora.
This sampling fills two requirements: firstly it ensures that each is representative of the entire \medline data set, secondly it preserves our ability to perform validation using the cut-year of 2014.
This allows us to identify connections that first occur in 2015 or later, which we will use to evaluate our network's performance.

We observe in our test corpora that \medline contains a significant number of single-sentence abstracts, typically just a title, that represent old documents that have not been entirely added.
For instance, the document with PMID 711285 consists of the single word ``hypertension.''
Additionally, \medline contains a number of non-English documents, such as PMID 21014169, which is in Spanish.
PMC, in contrast, contains a smaller set of more recent documents, which consist of fewer short or non-English abstracts.

\begin{table*}
\centering
\begin{tabular}{| c | r | r | r | c |}
\hline
Corpus ~&~ Total Words ~&~ Unique Words ~&~ Corpus Size ~&~ Median Words \\
 & & & & per Document \\
\hline
PMC Abstracts & 109,987,863 & 673,389 & 1,086,704 & 102 \\
PMC Full-Text & 1,860,907,606 & 6,548,236 & 1,086,704 & 1594\\
MEDLINE & 1,852,059,044 & 2,410,130 & 24,284,910 & 71 \\
1/2 MEDLINE & 923,679,660 & 1,505,672 & 12,142,455 & 71 \\
1/4 MEDLINE & 460,384,928 & 920,734 & 6,071,227 & 71 \\
1/8 MEDLINE & 229,452,214 & 565,270 & 3,035,613 & 71 \\
1/16 MEDLINE & 114,385,607 & 349,174 & 1,517,806 & 71 \\
\hline
\end{tabular}
\vspace{.5em}
\caption{\label{tab:corporaDetails}
The above table displays the corpus size for each experimental corpus we evaluated.
Note, each corpus has been filtered to only include documents available in XML and published before 2014.
Additionally, the above numbers represent each corpus after our initial text-cleaning process.
}
\end{table*}

\subsection{System Training and Query Process}

After selecting our corpora, we run the \emph{entire} \moliere network construction process, described in detail in~\cite{sybrandt2017moliere}, to create our knowledge network.
This process begins with phrase mining and \ftext~\cite{joulin2016fasttext}, a word embedding tool that allows us to numerically represent each n-gram in a dense, continuous, real-valued vector space.
For this paper, we chose an embedding dimensionality of $100$.
These n-gram embeddings allow us to project each document into the space as a centroid of its components.
From there, we create an approximate nearest-neighbors network for abstract centroids and n-grams (separately) using FLANN~\cite{muja_flann_2009}.
We join these layers with cross-cutting edges through TF-IDF.
Lastly, we introduce data from the Unified Medical Language System (\umls)~\cite{lindberg1993unified}.
This human-curated network represents ground-truth entities and connections that improve our network performance.
Then all link weights are renormalized. 
This entire process is automated by the source code available on-line\footnote{\url{http://github.com/jsybran/moliere}}.

We note that for validation purposes, we only include data published before 2015.
This means not only that we filter each corpora by publication date, but we also use the 2014 archival release of the \umls metathesaurus.
Additionally, by including the UMLS release to each corpus, we ensure that we are able to identify the needed entities for the later validation process.

To generate a hypothesis using our system, one supplies queries in terms of target words $a$ and $c$ (when performing a 1-to-1 query).
From there, the system identifies each in its internal knowledge network, and finds the shortest-path between the two.
We then extend shortest path  to include the \emph{cloud} of nearby documents by first finding the set of $p$ closest papers for each node along the shortest-path, and then taking the union of each set.
We then use PLDA+~\cite{liu2011plda+} to identify $k$ LDA topics within the extracted cloud, which we interpret as our hypothesis result.
For our tests here, we select $p=5000$ and $k=20$.

\subsection{Validation}

We evaluate each instance of \moliere using the technique established in~\cite{sybrandt2018validation}.
This process begins with a cut-year, which we chose to be 2014.
From there, we extract all \semmeddb predicates that were first published in 2015 or later~\cite{Kilicoglu2012}, and create a set of noise predicates through random sampling that have never been published.
This provides a set of positive (published) and negative (noise) hypotheses that our networks have not seen.

To evaluate the performance of \moliere, we generate both positive and negative hypotheses in order to evaluate the resulting topic models of each using a number of metrics.
Each metric captures a different relationship between $a$, $c$, and the resulting topic model.
These often include distances using the trained vector space, which makes the underling \ftext results incredibly important.
One of the metrics, \textsc{PolyMultiple}, is a polynomial combination of the others, with coefficients obtained through black-box optimization.
For the purposes of our tests here, we refit this metric for each system provided a set of 1 million training iterations. 

We then rank published and noise connections together with respect to each metric in order to create ROC plots.
We choose ROC curves as they have a direct relationship to ranking and because our validation method includes an equal number of positive and negative samples.
The area under each curve indicates an instance's ability to distinguish published connections from noise.
For a fair comparison between systems, we select a validation set of 2,000, equal parts published and noise, and use the same predicates on every system.
In addition to the qualitative result, we also measure memory, storage, and run-time requirements for each system.
All of our runtime measurements are run on 24 core machines with 126 Gb of memory, connected to a ZFS parallel file system.

There is a potential issue applying our validation scheme to full-text papers.
We get our predicates from \semmeddb, a data source that only extracts information from abstracts, and our validation makes the assumption that the published and noise sets are both unknown to the system under examination.
This implies that it could be possible for validation predicates to appear in full-text data that we do not intend, and there does not exist a reliable source of full-text predicates.
This stated, we note that authors typically attempt to highlight their key findings in their abstracts, and for a predicate to appear in our published validation set, its first occurrence  must date after 2014.
We find it unlikely that these new findings occur in any significant manner within the details of full-text papers, and by using \semmeddb as a standard, we are able to make better comparisons.

We additionally generate hypotheses regarding a recent highly-cited finding on every system in order to quantitatively evaluate each in a real-world use case.
The paper ``Mitochondrial Dynamics Controls T Cell Fate through Metabolic Programming'' (cited 131 times at time of writing) found in 2016 that the protein OPA1 is required for effector T-cells and not for memory T-cells.
We run two queries on our systems to relate OPA1 to immune effector cells and OPA1 to memory cells.

\section{Results}\label{sec:results}

\begin{figure*}[tb]
\centering
\includegraphics[width=0.22\linewidth]{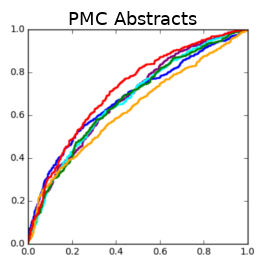}
\includegraphics[width=0.22\linewidth]{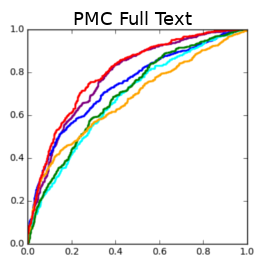}
\includegraphics[width=0.22\linewidth]{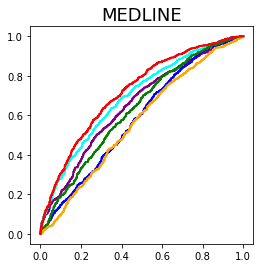}\\
\includegraphics[width=0.22\linewidth]{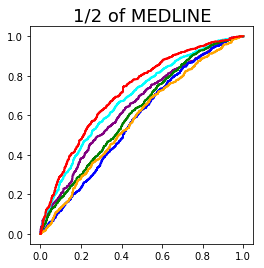}
\includegraphics[width=0.22\linewidth]{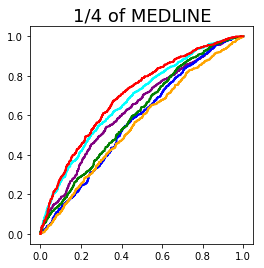}
\includegraphics[width=0.22\linewidth]{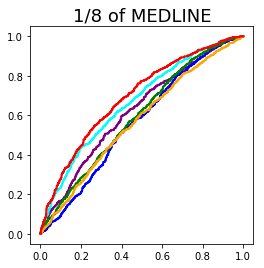}
\includegraphics[width=0.22\linewidth]{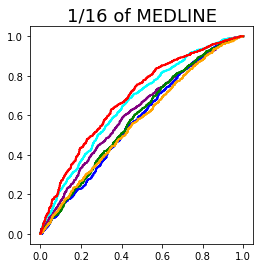}\\
\scriptsize
\centering
\begin{tabular}{|r | c c c c c c|}
\hline
&
{\footnotesize\color{purple} $\bullet$} $L_2$ &
{\footnotesize\color{blue} $\bullet$} \textsc{Centr}$L_2$ &
{\footnotesize\color{cyan} $\bullet$} \textsc{TopicPerWord} &
{\footnotesize\color{green} $\bullet$} \textsc{TopicCorr} &
{\footnotesize\color{orange} $\bullet$} \textsc{TopicWalkBtwn} &
{\footnotesize\color{red} $\bullet$} \textsc{PolyMulti} \\
\hline
PMC Ab.                                                                          
& 0.678
& 0.681
& 0.671
& 0.670
& 0.629
& 0.718 \\
% Hybrid
% & 0.657
% & 0.680
% & 0.669
% & 0.686
% & 0.622
% & 0.716 \\
PMC F.T.
& 0.777
& 0.738
& 0.680
& 0.696
& 0.674
& 0.795 \\
\medline
& 0.651
& 0.584
& 0.691
& 0.628
& 0.565
& 0.718 \\
1/2  \medline
& 0.643
& 0.578
& 0.684
& 0.615
& 0.580
& 0.717 \\
1/4  \medline
& 0.634
& 0.576
& 0.677
& 0.603
& 0.556
& 0.700 \\
1/8 \medline
& 0.621
& 0.566
& 0.666
& 0.593
& 0.570
& 0.691 \\
1/16 \medline
& 0.612
& 0.572
& 0.658
& 0.585
& 0.569
& 0.684 \\
\hline
\end{tabular}

\caption{\label{fig:res}
Above are the ROC curves for each experiment, accompanied by the AUC for key metrics, as described in~\cite{sybrandt2018validation}.
We evaluate a set of 2,000 predicates across each network to calculate each curve.
Note that the $L_2$ metric, which relies entirely on simple vector embeddings, is the best indication of embedding quality, while the \textsc{PolyMulti} metric combines others for peak performance.
}
\end{figure*}

After training instances of \moliere on each corpus and performing our validation task on each, we plot ROC curves for each across a number of metrics.
We summarize these results in Figure~\ref{fig:res} and discuss specific comparisons in the following sections.

% addresses q1 (do I need full text)
\subsection{PMC Abstracts vs Full-Text}
We see from Table~\ref{tab:corporaDetails} that the median PMC full-text contains almost $16\times$ as many words as the median PMC abstract.
For this reason, we expect that the resulting embedding space is of higher quality --- there is simply more training data.
We observe this when comparing the $L_2$ metric because this metric only evaluates hypothesis quality by taking the distance between $a$ and $c$, rather than incorporating topic model information.
The full-text $L_2$ area is 0.777 while the abstract $L_2$ result is 0.678.
This improvement is seen across many metrics, especially \textsc{PolyMultiple}, the trained polynomial combination of other metrics.
This is unsurprising because most metrics rely on the embedding space.

Looking practically we observe that constructing our full-text network takes $7\times$ the runtime, and twice as much storage.
Running each query takes $45\times$ longer (1h 15m for full-text vs. 1m 40s for abstracts), and substantially more memory (1.4Gb vs. 0.41Gb).
This is primarily due to the runtime of PLDA+, as it must read whole documents multiple times in order to fit a topic model.
Other differences in the query process come from network topology differences that result from the drastic change in document length.
Because we use TF-IDF to make cross-cutting connections between documents and keywords, we see that each document node has a substantially higher degree.

These network differences also account for qualitative differences in result quality between the PMC abstract and full-text systems.
The full-text system contains many more keywords that occur in practically every paper, such as gene, mouse, and cell.
While these words are certainly present in abstracts as well, their prevalence in methods and experimental sections \emph{biases them heavily in full text}.
Yet, we find their removal would substantially detract from our ability to interpret topics in the case of abstracts.

Looking at the best topics for the query between OPA1 and effector T-Cells, we see that the best topic for the abstract system has the leading words ``MGM1,'' ``mitochondrial,'' ``cell'' and ``GTPase'' while the full-text topic has the leading words ``cell,'' ``mitochondrion,'' ``mitochondrial,'' and ``protein.''
While both seem to capture the same content on a broad level, the abstract topic is much more focused on a single entity, MGM1 (which is a mitochondrial GTPase related to OPA1).
Still, neither network properly ranks effector T-Cells above memory T-Cells in relation to OPA1.

Overall, we see that abstracts give better qualitative and interpretable results using less time and memory, but full-text delivers a better vector space, and in turn allows for better evaluation via our metrics.
We anticipate that a hybrid approach that constructs the system with an embedding space trained on full-text but only abstract text for the purposes of running queries would be optimal.

% addresses Q2 (how many papers)
\subsection{MEDLINE Scaling Study}

Observing the results in Figure~\ref{fig:res}, we see the effect of adding additional papers of similar quality to our HG system.
Starting with the $1/16$ sample of \medline and working up to the whole data set, we see a consistent improvement across all results.
In a similar manner to the above, our metric increase seems to come primarily from an increase in embedding space quality.
We find that increasing our corpus size across the \medline experiments only has a marginal increase in $L_2$ performance, ranging from an ROC area of 0.604 in the $1/16$ sample to 0.638 in the $1/2$ sample.
Surprisingly, increasing the number of abstracts from the $1/2$ sample to the entirety of \medline has practically no effect on the resulting ROC curves.
We believe this is a side effect pertaining to the prevalence of very short \medline articles, as mentioned above.

Additionally, there is a discrepancy between our results here and those found in our previous work~\cite{sybrandt2018validation} wherein we achieved an ROC score of 0.833 on the same corpus.
In that case we created our network using an embedding space of dimensionality 500, as opposed to here where we use 100.
In that case, or $L_2$ metric was 0.783, which indicates that the higher dimensionality results in a significantly better embedding space.
In this study, we chose the smaller dimensionality to match the (typically) smaller corpus size.
Although further study is needed, we anticipate that given a higher vector dimensionality for these studies, we would see a greater difference between the \medline subsets as the higher dimensionality also implies it would be hard to train on smaller data sets.

In terms of performance, there is a bit of a difference in runtime between the scaled networks.
The $1/16$ system is able to run queries in about a minute and a half, using about 0.6 Gb of memory.
Meanwhile the $1/2$ system  requires about 3 minutes and 15 seconds, and 3 Gb of memory.
We note that the difference in runtime and memory usage primarily relates to the size of our overall network file (21 Gb for the $1/2$ and 3.2 Gb for the $1/16$ sample).
Our query algorithms rely on our parallel file system to help subset, load, and process this network in parallel, which helps keep our runtime down.
It is also worth noting that that the runtime of PLDA+ as well as our evaluation metrics is unchanged by the growing size of our network.
Each query still results in a similar number sized abstract cloud, which takes just as long to produce a topic model for.

Qualitatively, we also see a slight increase in result specificity as the corpus size increases.
This is not surprising as the $1/16$ sample likely excludes many important papers that would help to explain important connection.
In the case of OPA1 and effector T-Cells, as previously discussed, we see the $1/16$ best topic contains key words such as ``mitochondri,'' ``express,'' ``regul,'' ``activ,''  while the best topic for the $1/2$ system produces a best topic of ``protein,'' ``mitochondri,'' ``import,'' ``mitochondria.''
Although the $1/2$ sample is distinctly less informative from the PMC results above, its focus is much narrower than that of the $1/16$ sample.

\subsection{Cross-Comparison of Hypotheses using PMC and MEDLINE}

We observe in Table~\ref{tab:corporaDetails} that all of \medline contains approximately the same number of words as the PMC full-text dataset.
Additionally, we observe that the $1/16$ sample of \medline	 contains approximately the same number of documents as our PMC datasets.
Furthermore we note that the PMC abstract set is a subset of \medline.
This allows us to compare the two sources, and in doing so understand the effect document length, count, and quality have on our results.

We see that the PMC abstract set has approximately the same ROC area as the $1/2$ \medline set, yet has a similar number of words as the $1/16$ \medline set.
This shows us that a higher number of words per document is a big contributor to HG success.
Additionally, this increase in performance only increases the average runtime by ten seconds --- the PMC abstract system performs queries at about the same rate as the smallest \medline sample due to their similar network size.

% answers (probably move to conclusion?
We hypothesize that a future study wherein we only include \medline articles of sufficient length (at least two sentences) would be our best performer, or at least compete with the full-text results.
From our scaling tests we see that additional papers certainly improve results, and our cross comparison shows that a lower median document length is detrimental to HG results.
For these reasons a pruned set of \medline articles would have a larger median document length, but also contain an order of magnitude more documents than the PMC set alone.
A benefit of this strategy would likely be a substantially smaller runtime when compared to the full-text data set, as it would likely resemble the full \medline set in this aspect.

Looking quantitatively, comparing the OPA1---T-cell examples above, we note that sufficiently many abstracts provides better topics overall when compared to the full-text network.
This is because we see many frequent terms in full-text, such as ``fig,'' (meaning figure) that convey no additional content for our purposes.
We are reluctant, though, to expand our stopword list to include words like ``fig,'' or ``ref'' as their meanings in different contexts can be important, such as a fig-tree, 
or refer to the gene ``ALYREF.''
While additional development into entity extraction efforts could address these concerns directly, we note that the difference in author intent between abstract and full-text documents will still imply a difference in the sort of topics we will uncover.
\section{Tradeoffs}\label{sec:tradeoffs}

Interpreting our results from above, we can see a number of clear tradeoffs for HG when it comes to corpus quality.
The key issue is runtime vs. result quality, but other challenges such as data availability can also lead to tradeoffs beyond those captured in ROC curves.

We observe that longer documents require more processing --- they often have figures, extraneous text, and formulas, additionally, they often are distributed via PDF or were scanned through OCR.
These concerns do not even address legal challenges associated with collecting substantial collections of published papers.
As a result, uncovering their content for the purpose of HG is a non-trivial technical challenge that is likely to introduce noise.
We circumvented these issues by restricting our discussion to only papers available via XML, and we still faced many of these noise-related challenges.

Yet, our results indicate longer documents produce higher quality systems, assuming one can handle the drastically increased runtime.
This runtime limit becomes infeasible for many when considering large batch queries.
For instance, to find candidate genes related to a specific disease, we must run over 40,000 query pairs (one for each gene), and while these tasks are independent, we run into logistical challenges reserving the computational resources necessary to run 40,000 queries if each takes over an hour.

A more subtle tradeoff comes from our choice of topic models over other discovery methods.
Watson for Drug Discovery~\cite{spangler2014automated} uses many preprocessing techniques to make the runtime of individual queries much faster, which allow it to support full-text documents.
A potential downside to this is that users may want to run queries for entities that have not been identified.
While Watson is capable of handling these cases for many of its sub-systems, there are a number of query types that require preprocessed input.
\moliere's preprocessing involves identifying n-grams, which can pose similar problems, but then our approach compensates by facilitating queries between any pair of nodes, even papers and \umls terms.
But, as described in our results, the method introduces document length as a significant factor in our algorithmic complexity.
\section{Lessons Learned and Open Problems}\label{sec:lessions}

% We don't know the effect vector dimensionality has on our results
% we don't know how the phrase mninig has
% we see that preprocessing can alleviate runtime issues, new methods?
% new domains of interest!!!

\noindent{\bf Effect of Hyperparameters.}
Our network construction and validation method have a few hyperparameters that further study could help us inform their values.
In the network construction process, these include the dimensionality of our embedding space, the number of nearest-neighbors, and the weighting between layers of our network.
Following the example of Mikolov et al.~\cite{mikolov2013efficient,joulin2016fasttext}, we have previously used an embedding space of $500$, but in this work we reduce that to $100$.
We do so to see consistent performance across a number of corpus sizes, but we see reduced performance in our full-\medline experiment (when compared to our previous results in~\cite{sybrandt2018validation}).
Therefore, we note that larger corpus sizes ought to correlate with higher dimensionality vector spaces.
However, further study is necessary to understand the effect of our other network construction parameters.

In the query process, the two main hyperparameters are cloud size and number of topics.
We select a cloud of 5,000 documents per node along our shortest-path so that we have a sufficiently large sub-corpus even on paths with few hops.
Clearly there is a balance between quantity and quality --- the more documents we select, more of the documents will be unrelated to our query.
% Further study surrounding the relationship between each path-nodes' nearby document set in relation to its neighbors may provide some insight into automatically determining this parameter.
With regard to the number of topics, we find that fewer topics lead to more interpretable results, while more topics tend to aid in our automatic validation.
Yet, we have only experimented with topic values between 10 and 100 (20 for shown results).
%We plan to explore both heuristic methods to estimate topic count given an abstract cloud, as well as parameter-free models such as the hierarchical topic model.

\noindent{\bf Preprocessing.}
Automatic summarization and preprocessed topic models could each improve the runtime of full-text systems.
Automatic summarization should improve our signal-to-noise ratio and reduce the number of edges incident to each paper.
We could also generate a topic model for the paragraphs within each paper.
This may allow us to assemble a topic model for our abstract cloud by combining topics that are similar across multiple full-text documents.
This would be much faster than processing these documents directly for each query.

\noindent{\bf Domain Relevance.}
While we explore the effect of corpus size and document length in this work, we focus only on biomedical texts.
We hypothesize that we could improve \moliere results through the inclusion of additional documents from other fields such as physics, or more general sources such as Wikipedia.
These other sources would provide additional examples of technical writing to inform our embedding space, but may also implicitly diminish the effect non-medical terminology has in our network.
Because our cross-cutting edges between documents and keywords are weighted by TF-IDF, when we include documents from other fields, we expect that many methodology-related keywords that are shared between disciplines (such as ``study,'' and ``experiment'') will decrease in relevance as they are shared in practically every paper.
%At the same time, words that are unique to medicine that is unlikely to occur in this additional text will be proportionally less represented, thus increasing their TF-IDF importance.
This method may allow us to bias against phrases that many may consider stopwords, without explicit creating a stopword list, which as we mention above poses its own challenges.

\section{Deployment Challenges}\label{sec:challenges}
Parsing full-text papers is a major challenge for HG systems.
We found that only a subset of 1.7 million out of the 4.5 million documents in PubMed Central (PMC) was in XML format. 
Even then, some of the documents did not have an abstract (sometimes not the XML tag and sometimes not even the string "abstract" in the file).
Our strategy was chosen to remove poor quality XML parsings. 
We first parsed the XML of papers according to the listed specifications, but only successfully parsed 0.6 million documents. 
By looking at the patterns that were missed (e.g. tags nested in others like: [article meta] $\rightarrow$ [title group] $\rightarrow$ [article title]). 
We made our patterns less constraining to allow for more documents to be passed into the dataset. 
We ended up with 1.3 million documents with full-text and abstracts being parsed. 

When looking at some of the documents that failed, we found that they could have simply not tagged or failed XML parsers (e.g., with chains of emails or unrecognized symbols). 
A similar parsing task for abstracts was much simpler. As long as a paper had the abstract tagged, it was easy to add it to our dataset. 
However, we added abstracts if and only if the full-text was properly parsed.
%\subsection{Data Unification}

After stemming we were still left with the task of reducing the number of unique tokens during parsing. 
There was about 10 times the number of unique words within full-text documents than there were within the abstracts only. 
Even if we were comparing the larger set of abstracts to \medline, there were many fewer unique words (2.4M in all of \medline vs. 6.5M in PMC full-text). 
The number of unique words increased dramatically with every new addition of a particular document. 

%\subsection{Memory and Runtime issues}
The data storage requirements were highest, not at the end, but in the middle of the experiment when running AutoPhrase.
We started with 129 GB of textual data in XML format for 1.7 million documents from PMC (Downloaded in November 2017).
After text parsing and cleaning, we were left with 24 GB for full text and only 1.4 GB for abstracts.
Initially, we relied on \tmine for our phrase mining~\cite{el2014scalable} but had to switch after encountering memory and runtime problems. 
Although we were using computers with 500 GB of RAM, it was still not enough. 
We had to switch to a computer with 2 TB of RAM just to avoid crashing.
When finally getting \tmine to run properly, it was still too slow. 
Since our cluster only allows for jobs to run for 72 hours at a time, it was simply not able to finish running on the full-text dataset.

In order to mine phrases on full-texts we switched to AutoPhrase~\cite{shang2017automated}.
It was not only much faster, but also did not have the strict memory requirements that \tmine had. 
The information that was saved for the model was large (about 100 GB for the full text), but it was irrelevant since it was already using more than that in RAM. 
The process of topic modeling was simply memory intensive and unavoidable. 
Although we created the phrase model for the abstract dataset on a 24 core computer, it was still $4\times$ faster than the full-text phrase modeling on a 64 core computer. 
This amounted to an almost linear trend assuming processing scaled linearly. 
The ratio of runtime accounted for the number of cores, but not any other computer architecture specifics, ($4*64 / 24  \approx 10.7$) was almost equivalent to the difference in the data size ($24 \text{Gb} / 1.4 \text{Gb} \approx 17$).
\section{Conclusion}
\label{sec:conclusion}

%We evaluate our HG system, \moliere, on seven different corpora in order to answer three key questions.
In this paper, we systematically study the effects different types and sizes of data have on knowledge network based hypothesis generation systems. The experimental work is performed using \moliere \cite{sybrandt2017moliere} which extends traditional network-based HG systems using topic modeling and various hypothesis ranking techniques. The computational evaluation is demonstrated using seven different corpora in order to answer four key questions.

\noindent\textbf{What effect does corpus size and document length have on results?}
To answer this we compare the performance using ROC curves derived from our system when trained on PMC and \medline data sets.
We find that while increased corpus size does increase performance, document length is a better indication.
Our results show that selecting a corpus with a greater median document length of 30 words can have the same effect as selecting a corpus that is eight times larger.
However, we must emphasize that much longer documents  typically result in less interpretable (or too general) topic models of the hypotheses.

\noindent\textbf{How sensitive is a general-purpose HG system to hyperparameter value or input quality?}
\moliere is a general-purpose HG system that accepts queries for any terms covered in its input literature. Because of its scale, we anticipate that proper parameter tuning would require an analysis of more term pairs than is feasible.
This challenge is not present in more specialized systems, such as those targeting specific types of connections (gene-disease) or further specialized systems designed to explore a specific gene.
Still, we can explore our system's performance across a number of datasets given a fixed parameter setting in order to understand the hyperparamter's stability.
Our experiments show that a system trained on PMC abstracts outperforms one trained on a similar set of abstracts from \medline.
Looking at simple metrics, such as median document length and the length distributions, we observe a significant quality difference.
For this reason we conclude that quality is more important than quantity for HG.

\noindent\textbf{How many papers does a HG system need?}
We found that a set of at least 1-million papers is sufficient to achieve reasonable results, but more papers seem to improve result quality provided the underlying models are complex enough to capture the additional features.
These additional papers improve the performance of our embedding space, which underpins much of the \moliere query process, but this effect wanes if the dimensionality is too low.

\noindent\textbf{Are abstracts enough?}
We show that the tradeoff between quality and runtime is drastic when evaluating \moliere queries on full-text documents.
We compare our system trained on PMC abstracts against our system trained on the full-text versions of the same papers.
The longer documents cause longer topic modeling runtimes and result in a ROC area increase of 0.077 and a runtime increase from 100 seconds to 75 minutes.
While this tradeoff may be acceptable for some, we note that many batch query applications may not be able to afford this marginal improvement.

%\section*{Acknowledgment}

%more thanks here



\bibliographystyle{IEEEtran}
\bibliography{misc/bibFile,misc/alex}

\end{document}